# Improved Sputter Depth Resolution in Auger Thin Film Analysis Using In Situ Low Angle Cross-Sections

Uwe Scheithauer

SIEMENS AG, ZT MF 7, Otto-Hahn-Ring 6, 81730 München, Germany

Phone: + 49 89 636-44143, Fax: ...-42256, E-Mail: uwe.scheithauer@mchp.siemens.de[1]

## Abstract:

A none conventional approach for depth profiling of thin film systems with enhanced depth resolution has been developed using standard Auger microprobe instruments. For the preparation of an in situ low angle cross-section, the sample is partly covered by an appropriate mask. Utilising the edge of this mask, the sample is sputtered in the Auger microprobe with ions at nearly grazing incidence. In the shadow of the mask, this produces a low angle cross section through the thin film system. Then, a conventional depth profile is measured at the point of interest where part of the thin film system is covered only by a thin top layer. As demonstrated a considerable improvement of depth resolution $\Delta z/z$ can be obtained by this method.

---

[1] today (2015): uwe.scheithauer@siemens.com / scht.uhg@googlemail.com

## 1. Introduction:

In-depth profiling of thin film systems is one of the important applications of Auger electron spectroscopy. By this technique the sample surface is eroded by ion bombardment ("sputtering") and the residual surface is analysed after each sputter step. As a function of sputter time the depth distributions of the elements are recorded [1, 2]. In most cases the original in-depth distributions of the elements are modified by the sputtering process itself, because ion bombardment induces changes in the sample surface composition [3] and in the topography. Usually a rough sputter crater bottom develops, in particular for polycrystalline materials due to ion channelling. This is described by the depth resolution $\Delta z/z$, which represents the broadening of interfaces. To enhance depth resolution the sputter conditions can be optimised. In particular this is done by choosing the appropriate primary ion, reducing the primary ion energy, selecting a more grazing incidence ion impact angle [4, 5] or rotating the sample around its surface normal during sputtering to avoid permanent ion channelling on single grains [6, 7]. In literature even more complex approaches are reported [8, 9].

Here a new depth profiling technique utilising in situ low angle cross-sections is introduced. This approach is compared with more conventional methods of ion sputter depth profiling. The improvement of the depth resolution $\Delta z/z$ and the associated enhancement of detection limits at interface contaminations are demonstrated.

## 2. Experimental:

### 2.1 In Situ Low Angle Cross Section:

Fig. 1 illustrates the basic idea of an in situ low angle cross-section. The sample is partly covered by a mask. Utilising the edge of this mask in a first preparation step the sample is sputtered inside the Auger microprobe with ions at nearly grazing incidence. In the shadow of the mask this produces a cross-section through the thin film system with this low angle. For the in situ low angle cross-sections shown here an ion impact angle of 75° relative to the surface normal was used. Thus, the thickness of each layer is enlarged approximately by a factor of 3.9.[2] In a second step a conventional depth profile using standard sputter conditions is measured starting from a point where the interesting part of the thin film system is only covered by a thin top layer.

---

[2] see update at fig. 1

The layers can be identified by the SEM imaging and their compositions using Auger spectra. This layer identification and the choice of the appropriate starting point of the conventional depth profile may sometimes be time consuming. The surface of a low angle cross section in a polycrystalline material is not perfectly smooth but rugged to a certain extent (see fig. 2). In microscopic dimensions the gradient angle in some areas is smaller as given by the macroscopic parameters. Such an area is a good choice for measuring the depth profile, since on this area the contribution to depth resolution due to the lateral dimensions of the analysis area is reduced. At this moment it helps to have a good SEM resolution as provide by a field emission Auger microprobe.

## 2.2 Depth profile analysis:

The Auger depth profiles presented here were acquired using Physical Electronics PHI 660 and PHI 680 Auger microprobes. The PHI 660, an instrument with a $LaB_6$ electron emitter and a resolution of approximately 100 nm under analytical working conditions, is equipped with two differentially pumped ion guns (model 04-303). The ion guns are mounted with the same angle with respect to the primary electron beam and have a mutual azimuthal angular separation of 60°. The PHI 680 is an instrument with a hot field electron emitter and has an analytical resolution of approximately 30 nm. It is equipped with one differentially pumped ion gun (model 04-303). For all measurements the ion guns are operated with $Ar^+$ at 3 keV primary energy. For practical reasons the depth resolution is determined here using the sputter time difference of the 75% and 25% intensity values of the respective Auger signal.

# 3. Results:

## 3.1 Al on Si:

Fig. 2 shows Auger depth profiles through a 2.3 µm thick Al layer deposited on Si. The depth profiles were measured under different experimental conditions. In fig. 2 for each measurement the experimental parameters, in particular those of ion sputtering, the achieved depth resolution, the depth profile itself and a SEM image are shown row by row. The SEM images were recorded after the depth profile measurements. They show the surface topography that has developed during ion sputtering.

For the profile in the first row, sputter parameters typical for routine analysis in a surface analytical instrument were chosen. The attained depth resolution $\Delta z/z$ = 24.7 % is rather poor. The reason for this is obvious from the SEM image. Due to the different sputter rate on various grains a very rough surface has developed during ion bombardment. The analysis area is marked in the image. Areas of the same size were used for the other measurements shown here too, except the low angle cross-section. The area is large compered to the grain size of the analysed Al layer, therefore the Auger intensities measured during the depth profile represent the average over the different depths reached after a certain sputter time.

Using two ion guns instead of one improves the depth resolution to $\Delta z/z$ = 16.8 %. A pyramidal surface topography has developed during sputtering. A more grazing incident sputter ion beam is more effective in terms of depth resolution [4, 5]. Hence the depth resolution is improved to $\Delta z/z$ = 11.5 %. Note from the SEM image that the small cone structures point in the direction of the incident ion beam.

The main improvement is obtained by sample rotation [6, 7] and, even more, by a low angle cross section. Sample rotation during sputtering leads to a depth resolution of $\Delta z/z$ = 3.4%. The surface exhibits only small topographical features after sputtering. And using a low angle cross-section a depth resolution of $\Delta z/z$ = 1 % is finally obtained. During the low angle cross-section, a considerably rough surface develops. For the final depth profile the area has to be chosen very carefully. The measurement shown here was taken from the marked area located on the left hand side of the 10 μm marker in the high magnification SEM image.

In the sample rotation and low angle cross-section depth profile measurement, O is detected at the Al/Si interface. The integral of the quantified O signal can be normalised by the integral of the quantified Al signal. From the known thickness of the Al layer those of the Si oxide can be roughly estimated to be approximately 1.5 nm.

Low angle cross-sections made by mechanical polishing are proven not be useful here. Due to the water based polishing lubricant a thicker Al oxide layer had developed. Since the conventional Auger depth profile measurement has to start at a point where the original Al/Si interface is not disturbed by the oxide formation, an unsatisfied depth resolution was obtained due to an excess thickness of the Al oxide and Al layers that had to be removed.

### 3.2 Multilayer metallisation stack:

As a typical working example showing the improvement obtained by using a in situ low angle cross-section, the depth profile analysis of a multilayer metallisation stack is presented. The sample has the layer sequence Al, TiN, Ti, Ta, Cu, Ta and $SiO_2$ on a Si substrate. The Al layer is 800 nm thick. In fig. 3 a conventional sputter depth profile and a depth profile that has been measured utilising a low angle cross-section are compared. The conventional sputter depth profile was measured using 3 keV $Ar^+$ ions with an impact angle of 55° relative to the surface normal. For the sputter depth profile measured using the in situ low angle cross section 3 keV $Ar^+$ ions with an impact angle of 30° relative to the surface normal were used.

The depth resolution is improved by a factor of 8.2 for the in situ low angle cross section technique as extracted from the increase of the Cu signal. Due to this enhancement the layers and interfaces are much better resolved. At the Ti/Ta and the Ta/Cu interfaces of the upper Ta layer, O is detectable. Using internal reference spectra and non-linear least-square fit routines this O can be distinguished from the O in the $SiO_2$. In addition, from the in situ low angle cross section depth profile, it can be seen that no serious interdiffusion of the metal layers has occurred.

## 4. Conclusions:

With the in situ low angle cross-section sputter depth profiling method the depth resolution $\Delta z/z$ of sputter depth profiles can be considerably enhanced compared to other approaches. Utilisation of in situ low angle cross-section is the appropriate procedure to measure depth profiles of thin film systems that are covered by thicker polycrystalline layers. Interface contaminations as well as other features of analysed thin film systems are much easier to detect. In many cases an in situ low angle cross-section is the only possible method to gain this information.

## Acknowledgements:

For fruitful discussions and invaluable suggestions I would like to express my thanks to my colleagues, in particular to S. Ahmend and W. Hösler. The multilayer metallisation stack was prepared by H. Körner. This part of the work was sponsored by the BMBF project "FOKUM", project no. 0 1 M 2972 B.

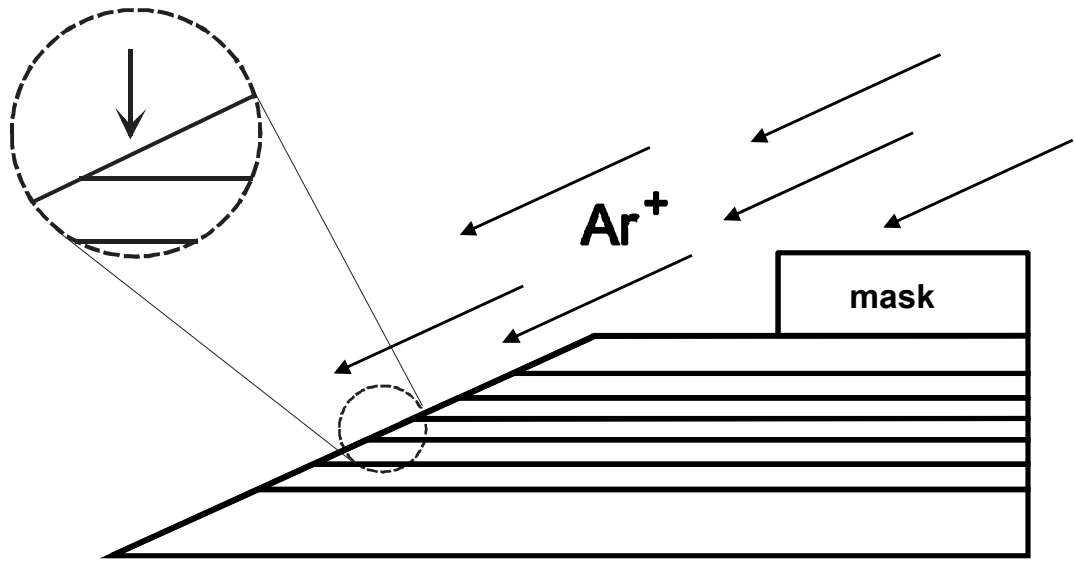

Fig. 1:  Basic schematic of an in situ low angle cross section. First the sample is sputtered with ions at nearly grazing incidence. Then a conventional depth profile is measured at the point of interest.

**Update:**

Later on it became clear, that the slope of the bevel, which is fabricated by the in situ low angle cross-sectioning, is flatter than given by the geometrical setup. For more details see:

U. Scheithauer, „In-Situ Low-Angle Cross-Sectioning: Bevel Slope Flattening due to Self-Alignment Effects", Appl. Surf. Sci., 255 (2009) 9062-9065,

DOI:10.1016/ j.apsusc.2009.06.103

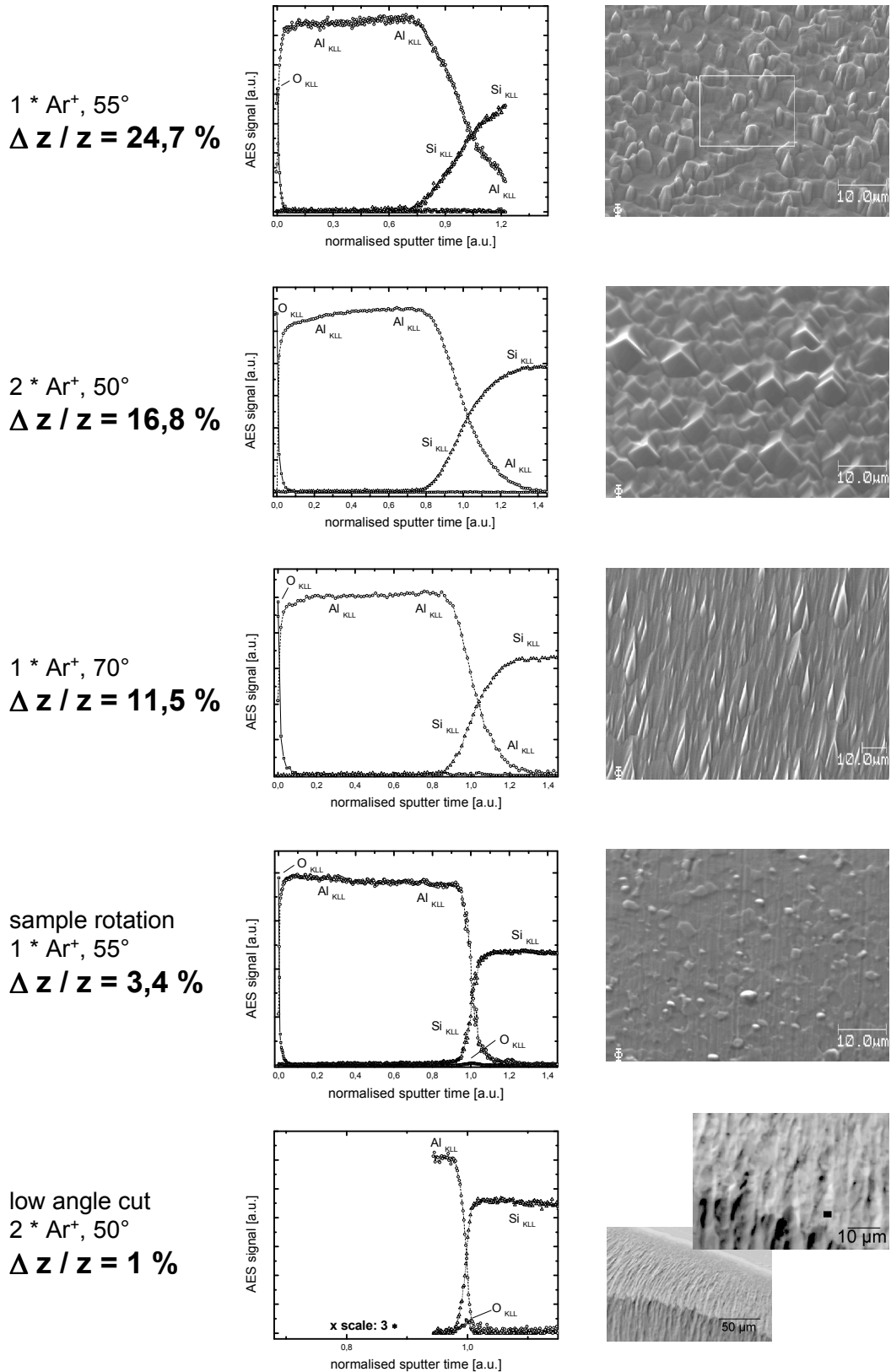

Fig. 2: Auger depth profile measurements through a 2.3 μm thick Al metallisation layer on Si, measured under different conditions. The experimental parameters and the attained depth resolution Δz/z are given. The SEM images were taken after the depth profile measurements, showing the topography of the sample surface that has developed during the ion sputtering.

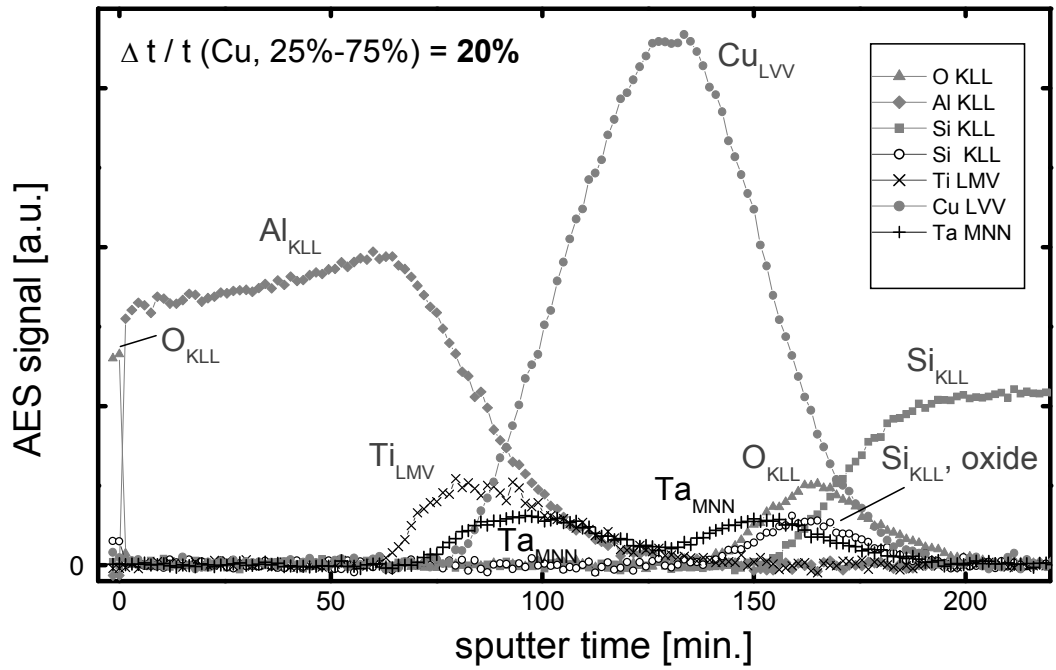

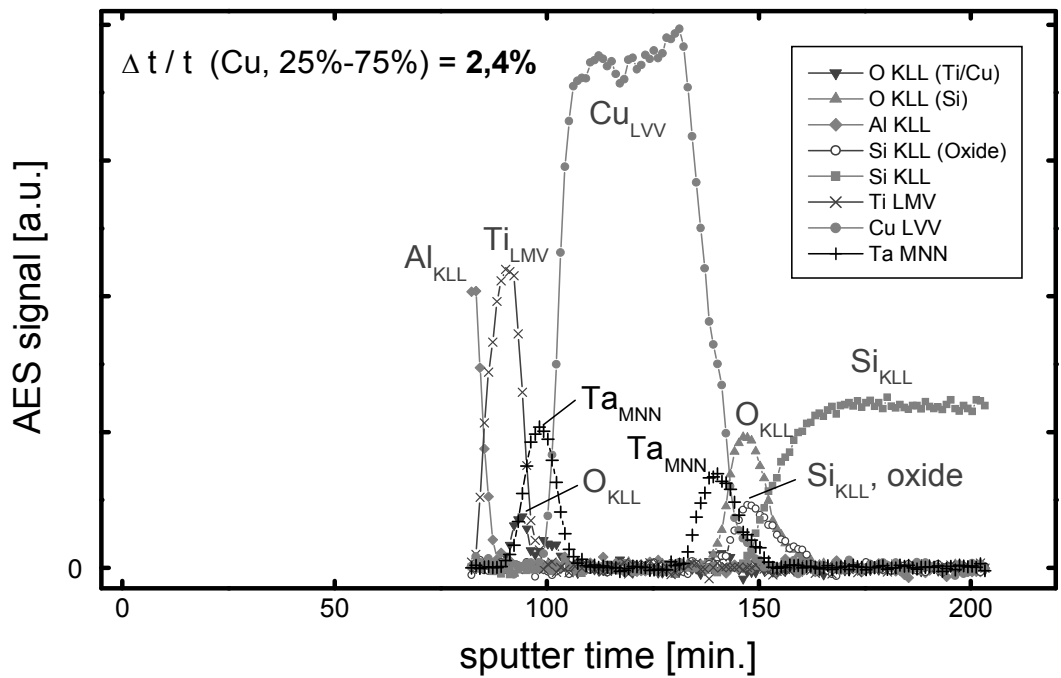

Fig. 3: Comparison of a conventional depth profile through a multilayer metallisation stack with a depth profile that has been measured utilising an in situ low angle cross section